\documentclass[aps,pre,showpacs,epsfig,preprint,amsmath,amssymb,subfigure,eqsecnum]{revtex4}
\usepackage{epsfig,amsmath,amssymb,bm,epsf,graphics,graphicx,psfrag,verbatim,subfigure,framed}
\usepackage{bbm}
\usepackage{mathrsfs}
\usepackage{amsfonts}
\usepackage{color}
\usepackage{dcolumn}

\newcommand{\rv}{{\vec r}}

\newcommand{\xv}{{\vec x}}

\newcommand{\zv}{{\vec z}}

\newcommand{\yv}{{\vec y}}
\newcommand{\kv}{{\vec k}}
\newcommand{\Tr}{{\rm Tr}}

\newcommand{\be}{\begin{equation}}
\newcommand{\ee}{\end{equation}}
\newcommand{\ba}{\begin{eqnarray}}
\newcommand{\ea}{\end{eqnarray}}
\begin{document}
\title{Particles inside Electrolytes with Ion-specific Interactions,  Their Effective Charge Distributions and Effective Interactions}
\author{Mingnan Ding, Yihao Liang, and Xiangjun Xing}
\affiliation{ Institute of Natural Sciences, and Department of Physics and Astronomy,
Shanghai Jiao Tong University,
Shanghai, 200240 China
 \\
    \email{dmnphy@sjtu.edu.cn,liangyihao@sjtu.edu.cn, xxing@sjtu.edu.cn.}         
}

\date{\today} 


\begin{abstract} 
In this work, we explore the statistical physics of colloidal particles that interact with electrolytes via ion-specific interactions.  Firstly we study particles interact weakly with electrolyte using linear response theory. We find that the mean potential around a particle is linearly determined by the {\em effective charge distribution} of the particle, which depends both on the bare charge distribution and on ion-specific interactions.  We also discuss the effective interaction between two such particles and show that, in far field regime, it is bilinear in the effective charge distributions of two particles.  We subsequently generalize the above results to the more complicated case where particles interact strongly with the electrolyte.   Our results indicate that in order to understand the statistical physics of non-dilute electrolytes, both ion-specific interactions and ionic correlations have to be addressed in a single unified and consistent framework. 
\end{abstract}

\maketitle

\section{Introduction}
The term ``ion-specific effects''~\cite{kunz-book-2010,Kunz-specific-ion-effects-colloids} usually refers to properties of electrolytes that depend on details of ions that are other than their electric charges.  In a broad sense, these details may include the effective sizes~\cite{Andelman-size} and polarizabilities \cite{Levin-polarizable} of hydrated ions, van der Waals interaction~\cite{Ninham-2001-Langmuir}, or other more complicated structures involving neighboring solvent molecules.  Because of their important implications in chemical and biological systems, study of ion-specific effects have attracted lots of attention in the past decade.  

There are different levels of modeling for relevant details of ions and solvent molecules.  At the microscopic level, one can carry out molecular dynamics (MD) simulations~\cite{Jungwirth-review} of electrolytes where both solvent molecules and ions are treated explicitly.  Such an approach can provide important clues about molecular mechanisms that are responsible for ion-specific phenomena, most notably the {\it Hofmeister series}.  At a more coarse-grained level, one can also incorporate relevant features (such as polarizability and dispersion forces) into the Poisson-Boltzmann theory, and obtain valuable insights.  For example,  Levin {\it et. al.}, used this method to explain Hofmeister series on surface tension of air/water interface and oil/water interfaces~\cite{Levin-polarizable}.  

In this work, we are more interested in the long scale consequences of ion-specific interactions. More specifically, we consider particles interacting with an electrolyte via a set of prescribed ion-specific interactions $\psi_{\mu}(\rv)$, beside conventional electrostatic interactions, and study ion densities and mean potentials around the particles, as well as the effective interaction between these particles.  An object that emerges from our analyses with central importance is the {\em effective charge distribution}, which depends  on both electrostatic and ion-specific interactions.  Another important concept is the {\em renormalized Green's function}.  We find that these two functions completely determine both the mean potential around a particle and the effective interaction between two particles.  Furthermore, for particles weakly interacting with electrolyte, we express the effective charge density in terms of the bare charge distribution and ion specific interactions.  Additionally we also relate the renormalized Green's function to various correlation functions of bulk electrolyte.




The remaining of this work is organized as follows. In Sec.~\ref{sec:LRT}, we use linear response theory to treat the simple case where the inserted particle(s) interacts weakly with electrolyte.  We  study the renormalized electrostatic Green's functions $G_{\! R}$ and relate it to various correlation functions.  More importantly, we define the {\em effective charge distribution} $\rho_{\rm eff}$ of a particle, and demonstrate how it completely determines the mean potential, as well as the effective interactions between particles.  In Sec.~\ref{sec:constituent-ions}, we define an {\rm effective charge distribution} $K_{\! \mu}$ for each specie of ion, and discuss its physical significance.  We also relate $K_{\! \mu}$ to the linear response properties of the electrolyte.  In Sec.~\ref{sec:PB}, we first show that the linear response equation we derived reduces to the linearized Poisson-Boltzmann theory when all correlation effects are ignored.  Additionally, using a PMF that takes into account correlations and ion-specific interactions, we derive a renormalized Poisson-Boltzmann equation.  In Sec.~\ref{sec:surfaces}, we generalize various results to the case where particles interact strongly with electrolyte.  Finally, we draw concluding remarks and envisage future directions. 

It is important to note that many of the results presented in this work were already derived by Kjellander and Mitchell (K\&M), in the setting of ``dressed-ion theory''~\cite{Kjellander:1992nr,Kjellander:1994xy,DIT-review-Kjellander}.   K\&M's theory is expressed using the formalism of density functional theory, is pivoted on a decomposition of the {\em direct correlation function} into short range part and a long range part, the latter {\em chosen} to be the Coulomb potential energy.  The summation of a long range function and a short range one is another long range function, hence such a decomposition is apparently non-unique.  By contrast, the formalism developed in this work is devoid of such a choice.  Additionally, we also treat the issue of ion-specific interactions and its interplay with electrostatic interactions more explicitly.  We shall explain further differences between our theory and the dressed-ion theory as appropriate along the way.

\section{Particles Weakly Interacting with Electrolytes}
\label{sec:LRT}
In this section, we shall treat the simple case where the inserted particle(s) can be treated as weak perturbations to the electrolyte, so that (static) linear response theory can be used to study both ion density profiles and mean potentials around the particles.  We shall derive the linear relations between external (both electrostatic and ion-specific) potentials and ion densities, from which the total mean potential $\phi$ can be calculated.  We shall then define the (renormalized) electrostatic Green's functions $G_{\! R}$, which relates the total mean potential $\phi$ to the {\em effective} charge density $\rho_{\rm eff}$, the latter being a linear superposition of charge density and ion-specific potentials.  Finally we shall show the effective interaction between two particles can be written as the sum of an electrostatic part and a remnant ion-specific part, with the latter decays faster than the former.   


\subsection{Free energy functional and density correlation functions}
\label{sec:free-energy}

Let the particle have a fixed charge density $\rho_{\rm ex}  (\rv)$, which generates a potential $\phi_{\rm ex}(\rv)$ via Coulomb's law:
\begin{subequations}
\label{rho-phi-ex}
\ba
\rho_{\rm ex}  (\rv) &=& - \epsilon \, \Delta  \phi_{\rm ex}(\rv),
\\
\hat{\rho}_{\rm ex} (\kv) &=&  \epsilon \, k^2  \hat{\phi}_{\rm ex}(\kv),
\ea
\end{subequations}
where $\Delta$ is the Laplacian.  Introduce the bare Green's operator: 
\be
G_{\! 0} = (- \epsilon \Delta)^{-1},
\ee
we can express the Coulomb's law Eqs.~(\ref{rho-phi-ex}) in the operator form:
\ba
\rho_{\rm ex} = G_{\! 0}^{-1} \phi_{\rm ex}, \quad
\phi_{\rm ex} = G_{\! 0} \rho_{\rm ex}. 
\label{rho-phi-ex-1}
\ea
A discussion of notations on operators and Fourier transforms used in this work is given in the appendix. 

We shall also assume that the particle interact additionally with all constituent ions via a set of ion-specific potentials  $\{ \psi_{\!\mu}(\rv), \mu = 1, \ldots, S \}$.  The total Hamiltonian of the perturbed electrolyte is then 
\ba
 H &=& H_0 + \sum_i \left[
q_i \phi_{\rm ex}(\xv_i) -  \psi_{\mu_i}(\xv_i) 
\right] 
\label{delta-H-app}
\ea
where $H_0$ is the Hamiltonian of the unperturbed, homogeneous electrolyte, whose concrete form does not concern us, whilst $\mu_i$ is the specie index of $i$-th ion.   Note that we have not included the self energies for $\phi^{\rm ex}$ and $\psi_{\! \mu}$ in the Hamiltonian.  We shall come back to this issue later.

It is convenient to define the ion number densities $n_{\mu}(\rv)$ and the charge density $\varrho (\rv)$ in the given micro-state, respectively:
\begin{subequations}
\ba
 n_{\mu}(\rv) &=& \sum_{i \in \mu} \delta(\rv- \xv_i), 
\label{rho_mu-def}\\
\varrho (\rv) &=& \sum_i q_i \, \delta(\rv- \xv_i)
= \sum_{\mu} q_{\mu}  n_{\mu}(\rv).  
\label{rho_q-def}
\ea
\end{subequations}
Note that in Eq.~(\ref{rho_mu-def}), the summation is over all ions belonging to specie $\mu$, whilst in Eq.~(\ref{rho_q-def}) the microscopic charge density $\varrho$ is a linear superposition of $n_{\mu}$.  
Note also that each ion carries a point-like charge in our theory, so that every ion contributes a delta function to $\varrho$.  
Using these relations, we can rewrite the Hamiltonian Eq.~(\ref{delta-H-app}) as
\begin{subequations}
\ba
H &=& H_0 + \int_{\rv} 
\left[ \varrho (\rv) \phi_{\rm ex}(\rv)
- \sum_{\mu}  n_{\mu}(\rv)
\psi_{\! \mu}(\rv)
\right], \quad  \\
&=& H_0 
-  \int_{\rv} \sum_{\mu}  n_{\mu}(\rv)
\left[  \psi_{\! \mu}(\rv)
- q_{\mu}  \phi_{\rm ex}(\rv) 
\right]  ,
\label{delta-H-app-2}
\ea
\end{subequations}

The free energy of the perturbed system is 
\ba
F[\phi_{\rm ex},\psi_{\! \mu}] &=& - T \log \Tr \, e^{-\beta H_0 
- \beta  \delta \! H}
\label{F-definition}
\\
&=& F_0 - T \log \left\langle 
 e^{  \beta  \int_{\rv} \sum_{\mu}  n_{\mu}(\rv)
\left[  \psi_{\! \mu}(\rv)
- q_{\mu}  \phi_{\rm ex}(\rv) \right] }
\right\rangle_0, 
\nonumber
\ea
where $\Tr$ means integration over  coordinates of all mobile ions, and 
\be
F_0 = - T \log \Tr  \, e^{-\beta H_0 }
\ee
is the free energy for the {\em homogeneous unperturbed electrolyte}, and $\langle \,\cdot \, \rangle_0$ means average over the Gibbs distribution $e^{-\beta H_0}$.  

Let $C_{\! \mu\nu}(\xv - \yv)$, $C_{\! \mu q}(\xv - \yv)$, and $C_{\! qq}(\xv - \yv)$ be the {\em connected} ion number-ion number, ion number-charge, and charge-charge correlation functions:
\begin{subequations}
\label{J-correlations}
\ba
C_{\! \mu\nu}(\xv - \yv)  & \equiv &  
\left\langle  n_{\mu}(\xv)   n_{\nu}(\yv) 
\right\rangle^c_0
=  \left\langle  n_{\mu}(\xv)   n_{\nu}(\yv) 
\right\rangle_0 - \bar{n}_{\mu} \bar{n}_{\nu} ,
\\
C_{\! \mu q}(\xv - \yv)  & \equiv &  
\left\langle  n_{\mu} (\xv) \varrho  (\yv) 
\right\rangle^c_0
= \sum_{\nu} q_{\nu} C_{\!\mu \nu }(\xv - \yv), 
\\
C_{\! qq}(\xv - \yv)  & \equiv &  
\left\langle \varrho (\xv)  \varrho (\yv) 
\right\rangle^c_0
= \sum_{\mu,\nu} q_{\mu} q_{\nu} 
C_{\!\mu \nu}(\xv - \yv). 
\ea
\end{subequations} 
Here $\bar{n}_{\mu}$ is the bulk ion number density of specie $\mu$, which satisfy the condition of overall charge neutrality:
\be
\sum_{\mu} q_{\mu}  \bar{n}_{\mu} = 0. 
\label{charge-neutrality}
\ee  
It is important to emphasize that  $C_{\! \mu\nu}$ defined here are {\em not} the {\em direct correlation functions} frequently used liquid stat physics.  The correlation functions $C_{\! \mu\nu}$ are related to {\em the pair correlation functions} $g_{\mu\nu}(\rv)$, {\em the total correlation functions} $h_{\! \mu\nu}(\rv)$ via
\ba
C_{\! \mu\nu}(\rv) &=&  \bar{n}_{\mu} \delta_{\! \mu\nu} \delta(\rv) 
+  \bar{n}_{\mu}\bar{n}_{\nu} 
\left( g_{\! \mu\nu}(\rv) - 1\right) 
\nonumber\\
&=&  \bar{n}_{\mu} \delta_{\! \mu\nu} \delta(\rv)
+ \bar{n}_{\mu}\bar{n}_{\nu} h_{\! \mu\nu}(\rv). 
\label{C-h-relation}
\ea
$g_{\mu\nu}(\rv)$ are also called the {\em radial distribution functions}.  For a discussion on various correlation functions frequently used in liquid state physics, see the classic textbook by Hansen and MacDonald \cite{Hansen:2013qd}. 

Let us define $\hat{C}_{\! \times \times}(\kv)$ as the Fourier transforms of ${C}_{\times \times}(\rv)$ (with $\times  = \mu, q$):
\ba
\hat{C}_{\! \times \times}(\kv) = \int_{\rv} {C}_{\! \times \times}(\rv) e^{i \kv \cdot \rv}. 
\ea
One can easily prove the following identities:
\begin{subequations}
\ba
\hat{C}_{\!\mu\nu}(\kv) (2 \pi)^3 \delta^3(\kv - \kv') 
&\equiv& 
\left\langle \hat{n}_{\! \mu}(\kv)  
\hat{n}_{\! \nu}(-\kv') \right\rangle^{\! c}_{\! 0},
\quad \quad \\
\hat{C}_{\!q\mu}(\kv) (2 \pi)^3 \delta^3(\kv - \kv') 
&\equiv& 
\left\langle \hat{\varrho}_q(\kv)  
\hat{n}_{\! \mu}(-\kv') \right\rangle^{\! c}_{\! 0},\\
\hat{C}_{\!qq}(\kv) (2 \pi)^3 \delta^3(\kv - \kv') 
&\equiv& 
\left\langle \hat{\varrho}_q(\kv)  
\hat{\varrho}_q(-\kv') \right\rangle^{\! c}_{\! 0}. 
 \ea
\end{subequations} 
Because of the linear relation between $\varrho$ and $n_{\mu}$, Eq.~(\ref{rho_q-def}), the following relations hold exactly:
\begin{subequations}
\label{C-relations}
\ba
\hat{C}_{\! q \mu}(\kv) &=&  \sum_{\nu} q_{\nu} \hat{C}_{\!\nu \mu}(\kv),
\label{C-relations-1}\\
\hat{C}_{\! qq}(\rv) &=&  \sum_{\mu} q_{\mu}  \hat{C}_{\!q \mu}(\kv)
= \sum_{\mu\nu}q_{\mu} q_{\nu} \hat{C}_{\!\nu \mu}(\kv). 
\label{C-relations-2}
\ea
\end{subequations}
As we shall show in this work, all linear response properties of the electrolyte can be characterized by these correlation functions.  

\subsection{Linear response equations}

As stated above, we assume that $\phi_{\rm ex}$ and $\psi_{\mu}$ are weak enough so that linear response theory is applicable.  It is then sufficient to expand the free energy Eq.~(\ref{F-definition}) up to the quadratic order in terms of $\phi_{\rm ex}(\rv)$ and $\psi_{\! \mu}(\rv)$.   The first order term in $\phi_{\rm ex}(\rv)$ vanishes identically, because $\langle \varrho (\rv)\rangle_0 = 0$, as dictated by charge neutrality and translational symmetry.  To the second order, we have  
\begin{subequations}
\label{F-quadratic-phi}
\ba
F[\phi_{\rm ex},\psi_{\! \mu}]   &=& F_0 
- \sum_{\mu} \bar{n}_{\mu} \int_{\rv} \psi_{\! \mu}(\rv)
\label{F-phi_ex-expansion-ENE}  \\
&-&  \frac{\beta}{2}  
 \sum_{\mu, \nu} \iint_{\xv, \yv}
\left[ \psi_{\! \mu}(\xv)
- q_{\mu}  \phi_{\rm ex}(\xv) 
\right] 
  C_{\! \mu \nu}(\xv - \yv)
\left[  \Psi_{\! \nu}(\yv)
- q_{\nu} \phi_{\rm ex}(\yv)
\right]
\nonumber\\
&=& F_0 - \sum_{\mu} \bar{n}_{\mu} \int_{\rv} \psi_{\! \mu}(\rv)
- \frac{1}{2}  \sum_{\mu, \nu} 
\left \langle \left( \psi_{\! \mu} - q_{\mu}  \phi_{\rm ex} \right)
\left| \beta C_{\! \mu\nu} \right|
\left(  \psi_{\! \nu} - q_{\nu} \phi_{\rm ex} \right) \right\rangle,
\nonumber
\ea
\end{subequations}
where in the last equality, we have switched to the Dirac brac-ket notations (introduced in the appendix, Eq.~(\ref{bracket-def})).  We shall not need higher order terms.  

Let us note that if we (formally) choose the ion-specific interactions such that $\psi_{\!\mu} = q_{\mu}  \phi_{\rm ex}$, it would exactly cancel the influence of the external electrostatic potential, and the perturbation to Hamiltonian Eq.~(\ref{delta-H-app}) would vanish identically.  Hence the perturbation of free energy must also vanish.  One can verify this explicitly by setting $\psi_{\!\mu} = q_{\mu}  \phi_{\rm ex}$ in Eq.~(\ref{F-phi_ex-expansion-ENE}).  The total free energy is however not invariant under this transformation, as the self-energies of $\phi_{\rm ex}$ and $\psi_{\!\mu}$ are generically different, and are not related by any simple transformation.


Let us now add to Eq.~(\ref{F-phi_ex-expansion-ENE}) the self-energy for $\phi_{\rm ex}$:
\ba
F_0 [\rho_{\rm ex}] &=&
 \frac{1}{2}
\left\langle \phi_{\rm ex} \! \left| 
G_{\! 0}^{-1} \right| \! \phi_{\rm ex} \right\rangle
=  \frac{1}{2}
\left\langle \rho_{\rm ex} 
\! \left| G_{\! 0} \right| \rho_{\rm ex} 
\! \right\rangle. 
\label{U-direct}
\ea
The self-energy of ion-specific interaction $\psi_{\! \mu}$ will not be discussed at this stage.  The change of total free energy due to $\phi^{\rm ex}$ and $\psi_{\! \mu}$ is then
\ba
\delta \! F^{\rm tot}[\phi_{\rm ex},\psi_{\! \mu}] &=&  
-  \sum_{\mu} \bar{n}_{\mu} \int_{\rv} \psi_{\! \mu}(\rv)
+ \frac{1}{2} \left \langle \phi_{\rm ex}\left| G_0^{-1}  \right| \phi_{\rm ex} \right\rangle
\nonumber\\
&-& \frac{1}{2}  \sum_{\mu, \nu} 
 \left \langle \left( \psi_{\! \mu} - q_{\mu}  \phi_{\rm ex} \right)
\left| \beta C_{\! \mu\nu} \right|
\left(  \psi_{\! \nu} - q_{\nu} \phi_{\rm ex} \right) \right\rangle, 
\label{F-phi_ex-expansion-2}
\ea

Taking the functional derivative of Eqs.~(\ref{F-definition}) and (\ref{F-phi_ex-expansion-2}) with respect to $\phi_{\rm ex}$, we obtain the {\em total average charge density}
\be
\rho^{\rm tot} = \frac{\delta F^{\rm tot}}{\delta \phi_{\rm ex}}
=  G_0^{-1}\phi_{\rm ex}  - \beta C_{\! qq} \phi_{\rm ex} 
+ \sum_{\mu} \beta C_{\! q\mu} \psi_{\!\mu}.
\ee
The first term in r.h.s is just the external charge density $\rho_{\rm ex}$, whereas the remaining two terms are due to  the mobile ions.  The total average potential can be obtained from $\rho^{\rm tot} $ via the Coulomb's law:
\ba
\phi &=& G_{\! 0} \rho^{\rm tot} 
= ( 1 - \beta G_{\! 0} C_{\! qq})  \phi_{\rm ex} 
 + \sum_{\mu} \beta G_{\! 0} C_{\! q \mu} \psi_{\!\mu} 
\nonumber\\
&=& (  G_{\! 0} - \beta G_{\! 0} C_{\! qq}  G_0) \rho_{\rm ex} 
 + \sum_{\mu} \beta G_{\! 0} C_{\! q \mu} \psi_{\!\mu}, 
\label{phi-rho_ex-psi}
\ea
where in the last step we have expressed $ \phi_{\rm ex} $ in terms of $\rho_{\rm ex} $ using Eq.~(\ref{rho-phi-ex-1}).  Products of operators are defined as convolutions in real space (see Eq.~(\ref{prod-def}) in the appendix) and as simple products in Fourier space.  Therefore all operators commute with each other.  Let us define the {\em renormalized Green's function} $G_{\! R}$ and {\em the effective charge density} $\rho_{\rm eff}$ via
\ba
G_{\! R} & \equiv & G_{\! 0} - \beta G_{\! 0} C_{\! qq} G_0, 
\label{phi-rho_ex-psi-2-1}
\\
\rho_{\rm eff} &\equiv& \rho_{\rm ex} 
+ \sum_{\mu} \beta G_{\! R}^{-1} G_{\! 0} C_{\! q\mu} \psi_{\!\mu}, 
\label{phi-rho_ex-psi-2-2}
\ea
 Eq.~(\ref{phi-rho_ex-psi}) can then be put into the following simple form:
\be
\phi = G_{\! R} \rho_{\rm eff}, \quad
G_{\! R}^{-1} \phi = \rho_{\rm eff}. 
\label{phi-rho_eff}
\ee

Let us define another kernel $\alpha$ via:
\be
\epsilon \, \alpha \equiv 
G_{\! R}^{-1} - G_0^{-1} 
= G_{\! R}^{-1}  
+ \epsilon \Delta.
 \label{G_R-G_0-def}
\ee
Using Eq.~(\ref{phi-rho_ex-psi-2-1}), we can further express $\alpha$ as
\ba
\alpha &=& \epsilon^{-1} \left( 1 - \beta G_{\! 0} 
C_{\! qq}\right)^{-1} \!\! \beta C_{\! qq}
=  \epsilon^{-1} \beta G_{\! R}^{-1} G_{\! 0}  C_{\! qq}. 
\label{alpha-def}
\ea
The Fourier space representation of Eq.~(\ref{alpha-def}) is (with help of Eq.~(\ref{phi-rho_ex-psi-2-1}))
\be
\hat{\alpha}(\kv) = 
 \frac{k^2 \beta \hat{C}_{\! qq}(\kv)}
 { \epsilon k^2 -  \beta \hat{C}_{\! qq}(\kv)}. 
\ee
Using Eq.~(\ref{G_R-G_0-def}), Eq.~(\ref{phi-rho_eff}) can be casted into the following form:
\begin{subequations}
\label{phi-alpha-LRT}
\be
\left( - \Delta + \alpha \right) \phi = \frac{1}{\epsilon} 
\rho_{\rm eff},
\ee
whose real space representation is a linear integro-differential equation for $\phi$:
\ba
- \Delta \phi(\xv) + \int_{\yv} \alpha(\xv - \yv)  \phi (\yv) &=& 
 \frac{1}{\epsilon}  \rho_{\rm eff} (\xv).  
\ea
\end{subequations}
This is the {\em linear response equation} that relates the mean potential to the effective charge density.  It differs from the well-known linear PB equation in two aspects: 1) a non-local kernel $\alpha$ due to long range electrostatic correlations, and 2) an effective charge density that takes into account ion-specific interactions.  If the ion-specific interactions $\psi_{\mu}$ are absent, Eqs.~(\ref{phi-alpha-LRT}) reduce to the linear response equation derived by Kjellander and Mitchell in the setting of dressed-ion theory \cite{Kjellander:1992nr,Kjellander:1994xy,DIT-review-Kjellander}.  

If we take the functional derivative of Eq.~(\ref{F-definition}) with respect to $\psi_{\!\mu}$ with $\phi_{\rm ex}$ fixed, we obtain the ion number density of specie $\mu$.  Taking the same derivative of Eq.~(\ref{F-phi_ex-expansion-2}), we find
\ba
\langle n_{\mu} \rangle = 
- \frac{\delta  F^{\rm tot}}{\delta \psi_{\!\mu}}
=  \bar{n}_{\mu} + \sum_{\nu} \beta C_{\! \mu\nu} 
\left( \psi_{\! \nu} - q_{\nu} \phi_{\rm ex} \right). 
\label{n_mu-F-0}
\ea
On the other hand, using Eqs.~(\ref{phi-rho_ex-psi}) and (\ref{rho-phi-ex-1}), we can express $\phi_{\rm ex}$ in terms of the total average potential $\phi$:
\be
\phi_{\rm ex}= G_{\! 0} G_{\!\! R}^{-1} \phi
- \sum_{\nu} \beta G_{\! 0} G_{\! R}^{-1}
G_{\! 0} C_{\! q\mu} \psi_{\!\mu}. 
\label{phi-phi_ex}
\ee
Further defining a set of kernels $\eta_{\mu\nu}$ via
\be
\eta_{\mu\nu} \equiv \beta C_{\!\mu\nu}
 + \beta^2 C_{\!q \mu} G_{\! 0} G_{\!\! R}^{-1}G_{\! 0} C_{\!q \nu}, 
  \label{eta-C-G}
\ee
and define $\delta \langle n_{\mu} \rangle = \langle n_{\mu} \rangle - \bar{n}_{\mu}$ as the deviation of ion number density, we can rewrite  Eq.~(\ref{n_mu-F-0}) as
\ba
\delta \langle n_{\mu} \rangle  
&=& - \beta  G_{\! R}^{-1} G_{\! 0} C_{\! q\mu}  \phi
 +  \sum_{\nu} \eta_{\mu\nu} \psi_{\! \nu}, 
 \label{n_mu-F-1}
\ea

In the next section, we shall derive a relation between the effective charge density $K_{\!\mu}$  of ion specie $\mu$ and correlation functions (c.f. Eq.~(\ref{K_mu-C}) ).  Anticipating this result, we can express Eqs.~(\ref{n_mu-F-1}) and (\ref{phi-rho_eff}) in the following form:
\begin{subequations}
\label{surface-structure-linear-0}
\ba
\delta \langle n_{\mu} \rangle  
&=& - \beta \bar{n}_{\mu}  
K_{\!\mu} \phi + \sum_{\nu} \eta_{\mu\nu} \psi_{\! \nu},
\label{n_mu-F} \\
\phi &=& G_{\! R} \, \rho_{\rm eff}
=  G_{\! R} \Big( 
\rho_{\rm ex} + \sum_{\mu} \beta \bar{n}_{\mu} K_{\! \mu} \psi_{\!\mu}
\Big). 
\label{phi-rho_ex-psi-2-3}
\ea
\end{subequations}
Furthermore, we shall also show in the next section that the kernels $\eta_{\mu\nu} $ decay faster than the renormalized Green's function $G_{\! R}$. Consequently, Eqs.~(\ref{surface-structure-linear-0}) show that the long scale physics is completely determined by two objects: the effective charge distribution $\rho_{\rm ex}$, which is a property of the particle, and the Green's function $G_{\! R}$, which is a property of the electrolyte. 



\subsection{An infinitesimally thin and permeable surface }
\label{sec:example}

Let us consider an infinitesimal thin, planar surface immersed in the electrolyte, with a bare surface charge density $\sigma_{\! 0}$, and a set of short range, ion-specific potentials $\{ \psi_{\! \mu}\} $.  The coordinate system is chosen such that the surface is at $x_3 = 0$, and the electrolyte fills the whole space.  We assume that $\sigma_{\! 0}$ and $\psi_{\! \mu}(\xv_{\perp})$ are weak enough to be treated as linear perturbations.  The effective charge density $\rho_{\rm eff}$ for the surface is then given by Eq.~(\ref{phi-rho_ex-psi-2-2}), with $\rho_{\rm ex}$ replaced by $\sigma_{\! 0} \, \delta(x_3)$.  If we further assume that $\psi_{\!\mu}(\yv)$ decays much faster than $K_{\! \mu}(\xv - \yv)$, we can make a further approximation: $\psi_{\!\mu}(\yv) = a_{\mu}\, \delta (y_3)$.  Hence $\rho_{\rm eff}$ reduces to
\begin{subequations}
\label{rho_eff-example}
\ba
\rho_{\rm eff}(\xv)  &=&  \sigma_{\! 0} \,  \delta(x_3)
+  \sum_{\mu} \beta \bar{n}_{\mu}  a_{\mu}
 \int_{\xv_{\perp}} \!\! \!\! K_{\! \mu}(\xv_{\! \perp}, x_3), \\
\hat{\rho}_{\rm eff}(\kv) &=&  \left[  \sigma_{\! 0} 
+  \sum_{\mu} \beta \bar{n}_{\mu}  a_{\mu}
\hat{K}_{\! \mu}(k_3)
\right] \delta^{(2)}(\kv_{\perp}). 
\ea
\end{subequations}

The mean potential can be obtained using Eq.~(\ref{phi-rho_ex-psi-2-3}).  We can write it explicitly as a Fourier transform:
\be
\phi(\rv) = \int_{\kv} \hat{G}_{\! R} (\kv ) 
\hat{\rho}_{\rm eff}(\kv)
e^{i \kv \cdot \rv}. 
\label{phi-rho-eff-example}
\ee
We shall further approximate the renormalized Green's function by its far field asymptotics:
\be
\hat{G}_{\! R} (\kv ) \approx \frac{1}
{\epsilon_{\! R} (k^2 + \kappa_{\! R}^2)},
\ee 
where $\kappa_{\! R}, \epsilon_{\! R}$ are respectively, the renormalized Debye length and renormalize dielectric constant.  Substituting this and Eq.~(\ref{rho_eff-example}) back into Eq.~(\ref{phi-rho-eff-example}), and carrying out the integral over $\kv$, we find that the far field asymptotics of the mean potential is given by
\ba
\phi(\rv) = \frac{\sigma_{\! R}}{2\, \epsilon_{\! R} 
\kappa_{\! R}} e^{-\kappa_{\! R} x_3},
\ea
where the parameter $\sigma_{\! R}$ should (obviously) be defined as the {\em renormalized surface charge density}, and is given by 
\ba
\sigma_{\! R} &=& \sigma_{\! 0} + \beta  \sum_{\mu}
 \bar{n}_{\mu} a_{\mu} \hat{K}_{\! \mu}(i \kappa_{\! R})
= \sigma_{\! 0} + \beta  \sum_{\mu}
 \bar{n}_{\mu}  a_{\mu} q_{\mu} ^R.
 \label{sigma_R-sigma_0}
\ea
Here $q_{\mu} ^R= \hat{K}_{\! \mu}(i \kappa_{\! R})$ is the renormalized charge of $\mu$ ion. \footnote{For a brief on the significance of renormalized charges, see reference \cite{DLLX-asym}. }  This result demonstrates that the renormalized surface charge density of a surface depends on two factors:  1) charge renormalization of constituent ions, and 2) ion-specific interactions between the surface and the constituent ions.  

\subsection{Effective interaction between two linear sources} 
Using the above results, the free energy Eq.~(\ref{F-phi_ex-expansion-2}) can also be rewritten as:
\ba
\delta\! F^{\rm tot} &=& - \sum_{\mu} \bar{n}_{\mu} \int_{\rv} \psi_{\! \mu}(\rv)
+ \frac{1}{2} \left \langle \rho_{\rm eff} \! \left| G_{\! R} \right|  \! \rho_{\rm eff} \right\rangle
- \frac{1}{2}  \sum_{\mu, \nu} 
\left\langle\psi_{\! \mu}  \! \left| \eta_{\mu\nu} \right| \! 
\psi_{\! \nu} \right\rangle,
\label{F-phi_ex-expansion-3}
\ea
Suppose we have two external sources $\{ \rho_{\rm ex}^A, \psi^A_{\!\mu}\}$ and $\{\rho_{\rm ex}^B, \psi^B_{\!\mu}\}$, which are well separated in space.  The total source is simply given by their superpositions:
\ba
\rho_{\rm ex} &=& \rho_{\rm ex}^{A} + \rho_{\rm ex}^{B}, 
\quad
\psi_{\!\mu} = \psi^A_{\!\mu} + \psi^B_{\!\mu}. 
\ea  
Substituting these back into the total free energy Eq.~(\ref{F-phi_ex-expansion-3}), and extracting the cross terms, we obtain the effective interaction between two linear sources:
\ba
\delta \! F^{\rm tot} &=&  
\left \langle \rho^{A}_{\rm eff} 
\! \left| G_{\! R}  \right| \! \rho^{B}_{\rm eff} \right\rangle
-  \sum_{\mu\nu} 
\left\langle\psi^A_{\mu} \! 
\left| \eta_{\mu\nu} \right| \! \psi^B_{\nu} \right\rangle
\label{Interaction-sources}
\\
&=&  \int_{\! \xv} \! \int_{\! \yv} 
\Big[
\rho^{A}_{\rm eff} (\xv)
G_{\! R}(\xv - \yv)\rho^{B}_{\rm eff}(\yv) 
- \sum_{\mu\nu} 
\psi^A_{\mu} (\xv) \eta_{\mu\nu}(\xv - \yv) \psi_{\nu}^{B}(\yv)
\Big]. 
\nonumber
\ea
where $\rho^{A}_{\rm eff}, \rho^{B}_{\rm eff}$ are defined by Eq.~(\ref{phi-rho_ex-psi-2-2}).   The first term is the  {\em renormalized electrostatic interaction}.  Note, however, the effective charge density  defined in Eq.~(\ref{phi-rho_ex-psi-2-2}) depends both on external charge distribution and on ion-specific interactions. The second term is remnant ion-specific interaction, which, as we shall show below, decays faster than the first term, and therefore can be neglected if the distance between $A, B$ is large (comparing with the Debye length).   Finally, we note that if there is a direct interaction between the ion-specific potentials $\psi^A$ and $\psi^B$, it should be added separately to Eq.~(\ref{Interaction-sources}).

\section{Effective Charge Distributions for Constituent Ions}
\label{sec:constituent-ions}

Let us fix a constituent ion of specie $\mu$ at the origin (in an otherwise homogeneous electrolyte), and measure the average ion number density $\langle n_{\nu} (\xv)\rangle_{(\mu, 0)}$ of specie $\nu$ at $\xv$ (including the fixed ion, if $\nu = \mu$) \footnote{The subscript $(\mu,0)$ means that there is an ion of specie $\mu$ fixed at the origin.}.  The interaction between $q_{\mu}$ and neighboring ions is usually strong and can not be described by linear response theory developed in the preceding section.  Nevertheless, we can always use Eq.~(\ref{phi-rho_eff}) to define an effective charge distribution $K_{\! \mu}$ for the ion $q_{\mu}$.  We shall show in this section that the set $\{K_{\! \mu}\}$ completely determines the linear response kernel $\alpha$.  Furthermore, $K_{\! \mu}$ also determines the effective interaction between the constituent ion and an externally imposed electrostatic potential.  Also, as a by-product, we shall show by a self-consistent argument that the kernels $\eta_{\mu\nu}$ are short ranged.

It is well known that $\langle n_{\nu} (\xv)\rangle_{(\mu, 0)}$  is related to the pair correlation function $g_{\mu\nu}$ via:
\ba
\langle n_{\nu} (\xv)\rangle_{(\mu, 0)} &=& 
 \delta_{\! \mu\nu} \delta(\xv)
 + \bar{n}_{\nu} g_{\mu\nu}(\xv)
\nonumber\\  
& =&  \frac{1}{\bar{n}_{\mu}} C_{\! \mu\nu}(\xv)
  +  \bar{n}_{\nu},
\label{n_mu-cond}
\ea
where in the second equality we have used the relation Eq.~(\ref{C-h-relation}).   The conditional average charge density is then given by
\ba
\langle \varrho (\xv)\rangle_{(\mu, 0)} 
= \sum_{\nu} 
q_{\nu }\langle n_{\nu} (\xv)\rangle_{(\mu, 0)}
= \frac{1}{\bar{n}_{\mu}} C_{\! q \mu}(\xv),
\ea
where in the last equality we have used Eq.~(\ref{n_mu-cond}) and Eq.~(\ref{charge-neutrality}).  Now the average potential $\phi_{\! \mu} (\xv)$ (due to both the fixed ion $q_{\mu}$ and other screening ions) can be obtained from $\langle \varrho (\xv)\rangle_{(\mu, 0)} $ via the Coulomb's law: 
\be
\phi_{\! \mu} = G_{\! 0} \langle \varrho \rangle_{(\mu, 0)} 
=  \frac{1}{\bar{n}_{\mu}} G_{\! 0} \, C_{\! q \mu}.  
\label{phi_mu-1}
\ee
Let us now define an {\em effective charge distribution} $K_{\! \mu}$ for the ion specie $\mu$, according to the linear response equation Eq.~(\ref{phi-rho_eff}):
\be 
K_{\! \mu} \equiv G_{\! R}^{-1}  \phi_{\!\mu} , \quad
\phi_{\! \mu} = G_{\! R} K_{\! \mu} .
\label{phi-rho_eff-mu}
\ee
Comparing this with Eq.~(\ref{phi_mu-1}), we obtain $K_{\! \mu}$ in terms of correlation functions:
\be
K_{\! \mu} = G_{\! R}^{-1}  \phi_{\!\mu}
=  \frac{1}{\bar{n}_{\mu}} 
G_{\! R}^{-1} G_{\! 0} \, C_{\! q \mu}.
\label{K_mu-C}
\ee
Using the exact relation (\ref{C-relations-2}) between $C_{\! qq}$ and $C_{\! q\mu}$, we can establish a similar relation between kernels $\alpha$ and $K_{\!\mu}$ (given respectively by Eqs.~(\ref{alpha-def}) and (\ref{K_mu-C})):
\be
\sum_{\mu} \frac{1}{\epsilon} \beta 
\bar{n}_{\mu} q_{\mu}  K_{\!\mu} 
= \alpha. 
\label{K-alpha-relation-2}
\ee
This relation was first established by Kjellander and Mitchel in 1990's. \cite{Kjellander:1992nr,Kjellander:1994xy,DIT-review-Kjellander} 

To compute the effective charge distributions $K_{\! \mu}$ for all constituent ions is clearly a difficult matter. Nevertheless, by comparing with the corresponding quantity Eq.~(\ref{phi-rho_ex-psi-2-2}) in the linear response theory, we can easily see that $K_{\! \mu}$ depends on both electrostatic correlations and ion-specific interactions between ions, and these dependences are generically nonlinear.  

\subsection{Potentials of mean force for constituent ions}
Assuming now that the system is perturbed by external potentials $\phi_{\rm ex}, \psi_{\!\mu}$.  The number density $\langle n_{\mu} \rangle$ is related to the potential of mean force (PMF) $U_{\! \mu}$ of ions of specie $\mu$ via the  Gbbs-Boltzmann distribution:
\be
\langle n_{\mu} \rangle 
= \bar{n}_{\mu} \, e^{-\beta U_{\! \mu} }.
\label{n_mu-F-2}
\ee
For weak external perturbations, $\beta U_{\! \mu}$ is small, so we can expand the exponential to the first order.   Comparing with Eq.~(\ref{n_mu-F}), we find that to the leading order, the PMF is a linear functional of $\phi_{\rm ex}, \psi_{\!\mu}$:
\be
U_{\! \mu} = K_{\!\mu} \phi - \frac{1}{\beta \bar{n}_{\mu}}
 \sum_{\nu} \eta_{\mu\nu}  \psi_{\! \nu}
 + O(\phi, \psi_{\! \mu})^2. 
 \label{U_mu}
\ee
Note, however, by keeping only linear terms, we are ignoring the polarizability of ions.  It is remarkable   that the effective charge distribution $K_{\! \mu}$ that generates the mean potential according to Eq.~(\ref{phi-rho_eff-mu}) is also responsible for the interaction between the ion and an externally imposed electrostatic potential.  



\subsection{$\eta_{\mu\nu}$ is short ranged}

It is well known that the pair correlation functions $g_{\! \mu\nu}$ are related to the two-ion PMFs $U_{\! \mu\nu}$ via $g_{\! \mu\nu}(\rv) = e^{- \beta U_{\! \mu\nu}(\rv)}$.  Usage of this in Eq.~(\ref{C-h-relation}) leads to 
\be
C_{\! \mu\nu}(\rv) = \bar{n}_{\mu} \bar{n}_{\nu} 
\left[ e^{-\beta U_{\! \mu\nu}(\rv)} - 1 \right]
 + \bar{n}_{\mu} \delta_{\! \mu\nu} \delta(\rv). 
 \label{C-identity-2}
\ee
On the other hand, Eq.~(\ref{eta-C-G}) can be rewritten into the following form:
\be
C_{\! \mu\nu} = - \beta \bar{n}_{\mu} \bar{n}_{\nu} K_{\!\mu} G_{\! R}K_{\! \nu}
+ \beta^{-1} \eta_{\mu \nu}. 
 \label{C-identity-3}
\ee
Comparing Eqs.~(\ref{C-identity-3}) with (\ref{C-identity-2}), we obtain the following expression for $\eta_{\mu\nu}$:
\ba
\beta^{-1} \eta_{\mu\nu} =  \bar{n}_{\mu} \delta_{\! \mu\nu} \delta(\rv)
+ \bar{n}_{\mu} \bar{n}_{\nu} 
\left[ e^{-\beta U_{\! \mu\nu}(\rv)} - 1
+  \beta  K_{\!\mu} G_{\! R}K_{\! \nu} \right]
 \label{C-identity-4}
\ea
The physical significance of $U_{\! \mu\nu}$ is {\em the effective interaction free energy between two ions}.  In the far field,  we expect that $U_{\! \mu\nu}$ is asymptotically given by the electrostatic interaction between their effective charge distributions $K_{\! \mu}$ and $K_{\! \nu}$, according to Eq.~(\ref{Interaction-sources}):
\be
U_{\! \mu\nu} = K_{\! \mu} G_{\! R}K_{\! \nu} 
+ \mbox{short ranged}, 
\label{U-farfield-1}
\ee
where ``short ranged'' denotes some function that decays faster than $G_{\! R}$.  Substituting this back into 
Eq.~(\ref{C-identity-4}), we conclude that $\frac{1}{\beta} \eta_{\mu\nu}$ must be short ranged, i.e., it must decay faster than $G_{\! R}$.  This in turn implies that in Eq.~(\ref{Interaction-sources}), the far field asymptotics of the effective interaction between two sources is controlled by the first term, a result that is consistent with Eq.~(\ref{U-farfield-1}).  

In summary, using all the ion-ion correlation functions $C_{\! \mu\nu}$, we can obtain the effective charge densities $K_{\! \mu}$ for all species of constituent ions, as well as the renormalized Green's function $G_{\! R}$ and the short range kernels $\eta_{\mu\nu}$.   In this sense, all linear response properties of an electrolyte are encoded in their  correlation functions.  All results discussed in this section have been obtained by Kjellander and Mitchell \cite{Kjellander:1992nr,Kjellander:1994xy,DIT-review-Kjellander} in the {\em dressed ion theory}.  Kjellander and Mitchell's original derivation is based on a separation of direct correlation functions into a long range part and a short range part, which seems to posses certain degree of arbitrariness.  As we have demonstrated in this section, the quantities $K_{\! \mu}$, $\phi_{\mu}$, $\eta_{\mu\nu}$, etc. are uniquely defined and therefore must be independent of arbitrary choices.   


\section{Poisson-Boltzmann Theory and Beyond}
\label{sec:PB}

Traditionally, Poisson-Boltzmann (PB) theory starts with the following approximation about the PMF of constituent ions:
\be
U^{\rm PB}_{\mu} = q_{\mu} \phi(\rv), 
\label{U-PB}
\ee
where $q_{\mu}$ is the bare charge, and $\phi(\rv)$ is the mean potential at $\rv$ {\em in the absence of the ion}.  Comparing this with Eq.~(\ref{U_mu}), we see that it is equivalent to approximating $K_{\! \mu}$ by the bare charge distribution:
\ba
K^{\rm PB}_{\! \mu}(\rv) = q_{\mu}  \delta(\rv).  
\label{K-PB}
\ea
{\em This amounts to ignoring all correlations as well as all possible ion-specific interaction between ions}.  Substituting this back into Eq.~(\ref{K-alpha-relation-2}), we find:
\begin{subequations}
\label{alpha-PB}
\ba
\alpha^{\rm PB}(\rv) &=& \kappa_0^2\, \delta(\rv), \\
\kappa_0^2 &=& \epsilon^{-1} \beta \sum_{\mu} 
\bar{n}_{\!\mu} q_{\mu} ^2. 
\ea
\end{subequations}
$\kappa_0$ is the inverse of {\em the bare  Debye length.}  The Green's function can be obtained using Eq.~(\ref{G_R-G_0-def}): 
\be
G_{\! R}^{\rm PB} = \frac{1}{\epsilon   \left( - \Delta + \kappa_0^2 \right)}.
\ee
The real space representations is the well known {\em screened Coulomb potential}:
\be
G_{\! R }^{\rm PB} (\rv)  = 
\frac{1} {4 \pi \epsilon r}
e^{-\kappa_0 r}. 
\label{G_R-PB}
\ee
Using Eqs.~(\ref{alpha-PB}) and (\ref{phi-rho_ex-psi-2-2}), (\ref{phi-alpha-LRT}) becomes the linearized PB equation:
\ba
&& \left( - \Delta + \kappa_0^2 \right) \phi 
 = \rho_{\rm ex} + \sum_{\mu} \beta \bar{n}_{\mu} q_{\mu} \psi_{\mu}, 
\ea
where $\rho_{\rm ex}$ and $\psi_{\mu}$ pertain to the {\em externally inserted particle}.  We can also use Eq.~(\ref{G_R-PB}) in Eq.~(\ref{Interaction-sources}) to write out the effective interaction between two charge distributions in the PB approximation:
\be
U_{\!A\! B} = \int_{\xv} \int_{\yv}
\rho^{A}_{\rm eff} (\xv)
\rho^{B}_{\rm eff}(\yv) 
 \frac{e^{-\kappa_0   |\xv - \yv| }} {4 \pi \epsilon |\xv - \yv|}. 
\ee
For charged hard sphere particles, one can easily show that the above result reduces to the well-known DLVO theory.  

Finally, substituting Eq.~(\ref{K-PB}) back into Eq.~(\ref{K_mu-C}), we find the correlation functions $\hat{C}_{\! q \mu}$ and $\hat{C}_{\! q q}$ in the framework of PB:
\begin{subequations}
\ba
\hat{C}_{\! q \mu}(\kv) &=& \frac{q_{\mu}  \bar{n}_{\mu} k^2}
{k^2 + \kappa_0^2}, \\
\hat{C}_{\! q q}(\kv) &=& \sum_{\mu} q_{\mu}  \hat{C}_{\! q \mu}(\kv)
= \frac{ \epsilon \kappa_0^2 k^2}
{ \beta (k^2 + \kappa_0^2)}. 
\ea
\end{subequations}
The PB theory does not say how we should deal with the short range part of the two-ion PMF Eq.~(\ref{U-farfield-1}). 

On the other hand, if we use the approximation Eq.~(\ref{U-PB}) in Eq.~(\ref{n_mu-F-2}) and further substitute the latter into the exact Poisson Equation:
\be
\label{Poisson-eq}
- \epsilon \nabla^2 \phi(\rv) 
= \rho^{\rm tot} 
= \rho_{\rm ex} + \sum_{\mu} q_{\mu} \langle n_{\mu}\rangle,
\ee
we obtain the well-known nonlinear Poisson-Boltzmann equation:
\be
- \epsilon \nabla^2 \phi(\rv) 
= \rho_{\rm ex} + \sum_{\mu} q_{\mu} 
\bar{n}_{\mu} e^{-\beta q_{\mu} \phi(\rv)}.  
\ee
The nonlinear PB equation suffers from the same weakness as the linearized PB equation, i.e., they both ignore the correlation effects between ions.  If we instead use Eq.~(\ref{U_mu}), which is correct up to the first order in $\phi, \psi_{\mu}$, in Eq.~(\ref{Poisson-eq}), we obtain
\be
- \epsilon \nabla^2 \phi(\rv) 
= \rho_{\rm ex} + \sum_{\mu} q_{\mu} 
\bar{n}_{\mu} e^{-\beta K_{\mu} \phi 
+ \bar{n}_{\mu}^{-1} \eta_{\mu\nu} \psi_{\nu} }.  
\label{RG-PBE}
\ee
 It can be called {\em the renormalized nonlinear Poisson-Boltzmann equation}, since it takes into account charge renormalization effects, as well as ion-specific interactions.  This equation, of course, is useful only if we know $K_{\mu}$ and $\eta_{\mu\nu}$, either approximately or exactly.  In reference \cite{DLLX-asym}, we use this equation to calculate the effective charge distributions $K_{\! \mu}$ of constituent ions in the asymmetric primitive model, where ion-specific interactions are absent. In the future, we shall use the same equation to study model electrolytes with ion-specific interactions.


\section{Particles Interacting Nonlinearly with Electrolytes}
\label{sec:surfaces}
In this section, we shall consider particles inside electrolytes that interact strongly with the electrolyte,  and hence can not be treated as small perturbations.   Nevertheless, we can always define an effective charge distribution  $K$ for the inserted particle, like what we have done for the constituent ions in Sec.~\ref{sec:constituent-ions}.   Using statistical mechanics, we shall show that the effective charge distributions also control the effective interaction between the particle and external potentials, as well as the effective interaction between two particles in the far field regime.  

 
Let $\Phi$ be the total mean potential, and $\delta \langle n_{\! \mu} \rangle $ the ion number densities, both of which are measurable experimentally, at least in principle.  
Following the original idea of Kjellander and Mitchell, we can use these quantities to {\em define} an {\em effective charge density} $K$ and {\em effective ion-specific potentials} $\Psi_{\!\mu}$ via the following relations:
\begin{subequations}
\label{eff-particles-dist}
\ba
K &\equiv& G_{\! R}^{-1} \Phi,
\label{n_mu-F-2-1}  \\
 \eta_{\mu\nu} \Psi_{\! \nu}
 &\equiv&  \delta \langle n_{\mu} \rangle 
 + \beta \bar{n}_{\mu}  K_{\! \mu} \Phi. 
\label{n_mu-F-2-2} 
\ea  
\end{subequations}
Turning these relation around, we obtain:
\begin{subequations}
\label{surface-structure-linear}
\ba
\Phi &=& G_{\! R}  \, K, 
 \\
\delta \langle n_{\! \mu} \rangle 
&=& - \beta \bar{n}_{\mu} 
 K_{\!\mu} \Phi + \eta_{\mu\nu} \Psi_{\! \nu},
\label{surface-structure-linear-3}
\ea
which are the analogue of Eqs.~(\ref{surface-structure-linear-0}).  Let us emphasize that $K$ and $ \{ \Psi_{\mu} \}$ are defined such that the linear response equations Eqs.~(\ref{surface-structure-linear-3}) hold exactly in the whole space, not just in the far field.    It is also useful to define another set of functions $ \{ J_{\!\mu} \}$ via
\ba
 J_{\!\mu} \equiv 
  \eta_{\mu\nu} \Psi_{\! \nu}
  = 
\delta \langle n_{\mu} \rangle 
 + \beta \bar{n}_{\mu}  K_{\! \mu} \Phi,
  \label{J_mu-def}
\ea
\end{subequations}
which contains essentially the same information as $\{ \psi_{\mu}\}$. 
 

\subsection{Statistical mechanical treatment}
Let us insert a particle into the electrolyte and further impose weak external potentials $\phi_{\rm ex}, \psi_{\!\mu}$.  The total Hamiltonian can then be written as
\be
H = H_0 + H_{\! E \! P} + H_{\! E \phi} + H_{\! P \phi}
+ H_{\! \phi\phi},
\ee
where $H_0$ is the Hamiltonian for the unperturbed electrolyte, $H_{\! EP}$ is the interaction between the electrolyte and the particle, which includes both electrostatic interaction and non-electrostatic interaction.  $H_{\! E \phi}$ is the interaction between the electrolyte and the externally imposed potentials $\phi_{\rm ex}, \psi_{\!\mu}$, whilst $H_{P\phi}$ is the interaction between the particle and the external potential, $H_{\phi\phi}$ the self-energy of $\phi_{\rm ex}$.  Only the last three terms depend on the external potentials, whose sum is 
\ba
H_{\! E\phi} + H_{\! P \phi} + H_{\! \phi\phi} &=& 
\int_{\rv}  [ \varrho (\rv) + \rho_{\! P} (\rv) + \frac{1}{2} \rho_{\rm ex} (\rv) ]
 \phi_{\rm ex}(\rv) 
\nonumber\\
&-& \int_{\rv} \sum_{\mu}  n_{\mu}(\rv) \psi_{\! \mu}(\rv). 
\ea
where $\rho_{\! P} (\rv)$ is the charge density due to the fixed particle.  Note that $\rho_{\rm ex} (\rv)$ and $\phi_{\rm ex}(\rv) $ are related to each other via Eqs.~(\ref{rho-phi-ex}).  Note also that we have assumed that the non-electrostatic potential $\psi_{\! \mu}$ does not directly interact with the inserted particle. \footnote{Such a term can always be added separately afterwards.} The total free energy is formally given by 
\ba
F[\phi_{\rm ex}, \psi] = - T \, \log \Tr \, e^{-\beta 
(H_0 + H_{\! EP} + H_{\! E\phi} + H_{\! P \phi}+ H_{\! \phi\phi})}. 
\ea
Now taking the functional derivative of both sides with respect to $\phi_{\rm ex}(\rv)$ and $\psi_{\! \mu}$, and
set $\phi^{\rm} = \psi_{\! \mu} = 0$, we cab obtain the following relations:
\ba
\left.\frac{\delta F}{\delta \phi_{\rm ex}(\rv)}\right|_0  &=& 
 \rho^{\rm tot}(\rv)
 = \left\langle \varrho (\rv) \right\rangle
  +  \rho_{\! P} (\rv) +  \rho_{\rm ex} (\rv) , \\
- \left.\frac{\delta F}{\delta \psi_{\! \mu}(\rv)} \right|_0 &=& 
\left\langle n_{\mu}(\rv) \right\rangle
= \bar{n}_{\mu} + \delta \!  \left\langle n_{\mu}(\rv) \right\rangle .
\ea
Note $ \rho^{\rm tot}(\rv)$ is the total average charge density.  Hence, if $\phi_{\rm ex}, \psi_{\!\mu}$ are small, they lead to the following first order correction to the free energy:
\ba
\delta F &=& \int_{\rv} \left[ 
 \rho^{\rm tot}(\rv) \phi_{\rm ex}(\rv)
- \delta \! \left\langle n_{\mu}(\rv) \right\rangle \psi_{\!  \mu}(\rv)
\right]
+ O(\phi_{\rm ex}, \psi_{\! \mu})^2,
\label{deltaF-1}
\ea
which can be understood as the {\em effective interaction between the particle and the externally imposed potentials}.   Note that $\langle n_{\mu}(\rv) \rangle$ and $\rho^{\rm tot}$ are define in the state with $\phi_{\rm ex}, \psi_{\!\mu}$ set to zero.

Now, let $\phi$ be the mean potential {\em in the absence of the particle}, but in the presence of the external potentials $\phi_{\rm ex}$. It is related to $\phi_{\rm ex}, \psi_{\! \mu}$ via Eq.~(\ref{phi-phi_ex}), which can be rewritten as (using Eq.~(\ref{K_mu-C}))
\ba
 \phi_{\rm ex} = G_0G_{\! R}^{-1} \phi
 - \beta \sum_{\mu} \bar{n}_{\mu}  G_{\! 0} K_{\! \mu} \psi_{\! \mu}.
 \label{phi_ex-phi}
\ea
Substituting this back into Eq.~(\ref{deltaF-1}), we obtain
\ba
\delta F &=& \int_{\rv} \left[ 
 \rho^{\rm tot} G_0G_{\! R}^{-1} \phi
- \left( \delta \! \left\langle n_{\mu} \right\rangle
+  \beta \bar{n}_{\mu} K_{\! \mu} \Phi
\right) \psi_{\! \mu}
\right], \nonumber \\
&=&  \left \langle  \rho^{\rm tot} |G_0G_{\! R}^{-1}  | \phi \right\rangle 
-  \sum_{\mu \nu} 
\left \langle \Psi_{\nu} | \eta_{\nu\mu}  |\psi_{\! \mu}
\right\rangle,
\ea
where in the second equality, we have used Eq.~(\ref{J_mu-def}).
Recall $G_0 \rho^{\rm tot}$ is the total mean potential $\Phi$ in the absence of $\rho_{\rm ex}, \psi_{\mu}$ (see the second paragraph of this section), hence according to Eq.~(\ref{n_mu-F-2-1}), $G_{\! R}^{-1} G_0 \rho^{\rm tot} = G_{\! R}^{-1} \Phi = K$ is the effective charge density of the inserted particle.  Therefore we can write the preceding equation in the following form:
\ba
\delta F = \left \langle K| \phi \right\rangle 
-  \sum_{\mu \nu} 
\left \langle \Psi_{\nu} | \eta_{\nu\mu}  |\psi_{\! \mu}
\right\rangle. 
\label{deltaF-2}
\ea
 Eq.~(\ref{deltaF-2}) is the general form of the PMF of the particle inside weak perturbations $\phi, \psi_{\mu}$.  Obviously, this is the analogue of Eq.~(\ref{U_mu}) for an external inserted particle. We re-emphasize that $\phi, \psi_{\mu}$ are the external electrostatic and non-electrostatic potential {\em in the absence of the particle}. 
 
\subsection{Effective interaction between surfaces}
Let us now consider two particles, labeled by $A, B$ respectively, inserted into the electrolyte.  In the absence of the other, particle $A/B$ generates a mean potential $\Phi^{A},\Phi^{B}$, as well as ion number density profiles $\delta^{A} \langle n_{\mu} \rangle,\delta^{B} \langle n_{\mu} \rangle$, respectively.  From these, we define the corresponding effective charge densities and ion-specific potentials via Eqs.~(\ref{eff-particles-dist}) and (\ref{J_mu-def}), which we rewrite below
\begin{subequations}
\label{eff-particles-dist-2}
\ba
K^{\! A} &=& G_{\! R}^{-1} \Phi^{ A},
 \\
J_{\mu}^{\! A} = \eta_{\mu\nu} \Psi_{\! \nu}^{\! A}
 &=&  \delta^A \langle n_{\mu} \rangle 
 + \beta \bar{n}_{\mu}  K_{\! \mu} \Phi^A, 
\\
K^{\! B} &=& G_{\! R}^{-1} \Phi^{ B},
 \\
J_{\mu}^{\! B} = \eta_{\mu\nu} \Psi_{\! \nu}^{\! B}
 &=&  \delta^B \langle n_{\mu} \rangle 
 + \beta \bar{n}_{\mu}  K_{\! \mu} \Phi^B. 
\ea  
\end{subequations}
We shall assume that the distance between two particles is large so that their mutual influences can be treated as linear perturbation.  Furthermore, we shall also assume that the number densities of all species of ions near a particle are not disturbed by the other particle.  In general, this kind of disturbances do appear, but is doubly screened, similar to all image charge effects.  These two assumptions are also made in the classical DLVO theory.  

Now the effective interaction between two particles can be obtained using Eq.~(\ref{deltaF-2}), by considering the particle $B$ as a {\em source of linear perturbation} to particle $A$.  Hence in r.h.s. of Eq.~(\ref{deltaF-2}) we only have to replace $K, J_{\! \mu}$ by $J^A, J_{\! \mu}^A$ respectively, and replace $\phi, \psi_{\!\mu}$ by $\Phi^B, \psi_{\!\mu}^B$ respectively.  This leads to 
\ba
U_{\! AB} &=& \left \langle K^{\!A} |
 G_{\! R} |K^{\!B} \right\rangle 
- \sum_{\mu \nu} 
\left \langle J^A_{\mu} |
\eta^{-1}_{\mu\nu} | J^B_{\nu}
\right\rangle
\nonumber\\
&=& \left \langle K^{\!A} | G_{\! R} | K^{\!B} \right\rangle 
- \sum_{\mu \nu}  \left \langle \Psi^A_{\!\mu}  | 
\eta_{ \mu\nu} | \Psi^B_{\!\nu}  \right\rangle. 
\label{Interaction-surfaces}
\ea
Even though this result appears identical to Eq.~(\ref{Interaction-sources}), it is important to note that Eq.~(\ref{Interaction-sources}) is applicable only to linear sources, whereas Eq.~(\ref{Interaction-surfaces}) is applicable to arbitrary surfaces, as long as they are widely separated in space.  It is important to note, however, $K^{A/B}$ etc. are defined as the effective charge distributions of particles $A/B$ in the absence of the other particle. Hence Eq.~(\ref{Interaction-surfaces}) ignores the polarization effect where proximity of the other particle changes the effective charge distribution of the first particles.  Finally we note that Eq.~(\ref{Interaction-surfaces}) looks similar to the one obtained by Kjellaner in reference \cite{DIT-review-Kjellander}.  The derivation presented here has the merit of disentangling properties of the bulk electrolyte from those of the inserted surfaces. It also makes clear that the effective interaction between two particles is predominantly electrostatic in the long scale, even though the effective charge distributions are generically renormalized by ion-specific interactions.  


Eqs.~(\ref{Interaction-surfaces}) and (\ref{deltaF-2}) are the main reason to define the effective charge distribution $K$ and the effective ion-specific potentials $\Psi_{\mu}$: They control the effective interaction between a particle and an external imposed linear sources, as well as the effective interaction between two particles. 

\section{Conclusion and Acknowledgement} 
We have developed a unified theoretical formalism for the statistical physics of non-dilute electrolytes with both electrostatic and non-electrostatic interactions, and have found that the long scale consequences of ion-specific interactions can be understood in terms of effective charge distributions.  In the future, we plan to compute the effective charge distributions for some concrete model systems with both electrostatic and ion-specific interactions.  

We thank NSFC (Grants No. 11174196 and 91130012) for financial support.  We also thank  Wei Cai for interesting discussions.  

\begin{thebibliography}{10}



\bibitem{kunz-book-2010}
Kunz, Werner, ed. 
\newblock Specific ion effects. 
\newblock Vol. 325. Singapore: World Scientific, 2010.


\bibitem{Kunz-specific-ion-effects-colloids}
Werner Kunz.
\newblock Specific ion effects in colloidal and biological systems.
\newblock {\em Current Opinion in Colloid and Interface Science}, 15:34--39,
  2010.

\bibitem{Andelman-size}
Borukhov, Itamar, David Andelman, and Henri Orland. 
\newblock "Steric effects in electrolytes: A modified Poisson-Boltzmann equation." 
\newblock Physical review letters 79.3 (1997): 435.

\bibitem{Levin-polarizable}
Levin, Yan. 
\newblock "Polarizable ions at interfaces." 
\newblock Physical review letters 102.14 (2009): 147803.\\
Levin, Yan, Alexandre P. Dos Santos, and Alexandre Diehl. 
\newblock "Ions at the air-water interface: an end to a hundred-year-old mystery?." 
\newblock Physical review letters 103.25 (2009): 257802.

\bibitem{Ninham-2001-Langmuir}
Boström, Mathias, David RM Williams, and Barry W. Ninham. 
\newblock "Surface tension of electrolytes: specific ion effects explained by dispersion forces." 
\newblock Langmuir 17.15 (2001): 4475-4478.
 
\bibitem{Jungwirth-review}
Jungwirth, Pavel, and Douglas J. Tobias. 
\newblock "Specific ion effects at the air/water interface." 
\newblock Chemical reviews 106.4 (2006): 1259-1281.

\bibitem{Kjellander:1992nr}
Roland Kjellander and D~John Mitchell.
\newblock An exact but linear and poisson---boltzmann-like theory for
  electrolytes and colloid dispersions in the primitive model.
\newblock {\em Chemical physics letters}, 200(1):76--82, 1992.

\bibitem{Kjellander:1994xy}
Roland Kjellander and D.~John Mitchell.
\newblock Dressed ion theory for electrolyte solutions: A Debye-H{\"u}ckel-like reformulation of the exact theory for the primitive
  model.
\newblock {\em The Journal of Chemical Physics}, 101(1):603--626, 1994.

\bibitem{DIT-review-Kjellander} 
Roland Kjellander.
\newblock ``Distribution Function Theory of Electrolytes and Electrical Double Layers: Charge Renormalisation and Dressed Ion Theory'' 
\newblock in Electrostatic Effects in Soft Matter and Biophysics (C. Holm, P. Kékicheff and R Podgornik, Eds; 
\newblock NATO Science Series, Kluwer Academic Publishers, Dordrecht 2001) pp. 317 – 364.

\bibitem{Hansen:2013qd}
Jean-Pierre Hansen and Ian R. McDonald.
\newblock {\em Theory of simple liquids: with applications to soft matter}.
\newblock Academic Press, 2013.


\bibitem{DLLX-asym}
Mingnan Ding, Yihao Liang, Bing-Sui Lu, and Xiangjun Xing.
\newblock Charge Renormalization and Charge Oscillation in Asymmetric Primitive Model. 
\newblock submitted to {\em Journal of Statistical Physics}. 

\end{thebibliography}

\appendix

\section{Notations: Fourier transform and Operator Formalism}
\label{sec:Notation}
In this section, we introduce some notations and identities about Fourier transforms and convolutions that will be frequently used in this work.  We shall use Fourier transform extensively in this section.  For a function $f(\rv)$ in real space, we shall use $\hat{f}(\kv)$ for its Fourier transform, which is related to $f(\rv)$ via
\begin{subequations}
\ba
\hat{f}(\kv) &=& \int d^3 \rv \, f(\rv) \, e^{ - i \kv \cdot \rv}
\equiv \int_{\rv}f(\rv) \, e^{ - i \kv \cdot \rv} ,
\\
f(\rv) &=& \int \frac{d^3 \kv}{(2 \pi)^3} \, \hat{f}(\kv) \, e^{ i \kv \cdot \rv}
\equiv \int_{\kv} \hat{f}(\kv) \, e^{ i \kv \cdot \rv}.
\ea
\end{subequations}
One can easily see that if $f(\rv)$ is both real and symmetric, so is its Fourier transform $\hat{f}(\kv)$:
\ba
&& f(\rv) = \overline{f(\rv)} = f(-\rv) = \overline{f(-\rv)}
\nonumber\\
   &\longleftrightarrow \,\,&
 \hat{f}(\kv) = \overline{\hat{f}(\kv)} = \hat{f}(-\kv)
 = \overline{\hat{f}(- \kv)} . 
\ea
Furthermore, if $f(\rv)$ decays to zero sufficiently fast as $\rv \rightarrow \infty$, then all moments of $f(\rv)$ are finite.  Of course, all odd order moments vanish identically because of symmetry.  Consequently, $\hat{f}(\kv)$ is an analytic function $k^2 = \kv\cdot\kv$.  All functions discussed in this work have these properties. 

We shall also frequently use convolutions:
\ba
C*g(\xv) \equiv \int_{\yv} C(\xv - \yv) f(\yv). 
\ea  
It is convenient to treat the first function $C(\xv)$ as an operator acting on the second function $f(\xv)$.  In this way, every function can be understood as operator, and all operators commute (because convolutions are commutative).  More specifically, we shall use  the following shorthand notations:
\ba
 f  &\equiv& f(\xv), \\
 C  f  &\equiv&  \int_{\yv} C(\xv - \yv) f(\yv) = C*f(\xv), \\
\langle f |C| g \rangle &\equiv& 
\iint_{\xv, \yv} f(\xv) C(\xv - \yv) g(\yv).  
\label{bracket-def}
\ea
Taking functional derivative of the last equality with respect to $g$, we find
\be
\frac{\delta }{ \delta \! f} \langle f|C| g \rangle
= C g,
\ee
which is a succinct form of the familiar identity:
\be
\frac{\delta}{\delta \! f(\xv)} 
\iint_{\xv, \yv} f(\xv) C(\xv - \yv) g(\yv)
= \int_{\yv} C(\xv - \yv) g(\yv). 
\ee
All functions (in real space) we use in this work are real valued, hence we do not need to worry about Hermitian conjugation.  Finally products of operators are also defined as convolutions:
\be
C D = \int_{\zv} C(\xv - \zv) D(\zv - \yv).
\label{prod-def}
\ee

\end{document}